\documentclass{rspublic}

\usepackage[dvips]{graphicx}
\usepackage{amssymb,amsfonts,amsmath}

\def\lsim{\mathrel{\rlap{\lower4pt\hbox{\hskip1pt$\sim$}}
    \raise1pt\hbox{$<$}}}                % less than or approx. symbol
\def\gsim{\mathrel{\rlap{\lower4pt\hbox{\hskip1pt$\sim$}}
    \raise1pt\hbox{$>$}}}                % greater than or approx. symbol
    
\def\be{\begin{equation}}    
\def\ee{\end{equation}}

\begin{document}

\title[Gravitational Wave Astronomy]{Gravitational Wave Astronomy:\\ Needle in a Haystack}

\author[N.J.~Cornish]{Neil J.~Cornish$^1$}

\affiliation{$^1$ Department of Physics, Montana State University, Bozeman, MT, 59717 USA}

\label{firstpage}

\maketitle

\begin{abstract}{Gravitational Waves, Bayesian Inference}

A world-wide array of highly sensitive interferometers stands poised to usher in a new era in
astronomy with the first direct detection of gravitational waves. The data from these instruments
will provide a unique perspective on extreme astrophysical phenomena such as neutron stars and
black holes, and will allow us to test Einstein's theory of gravity in
the strong field, dynamical regime. To fully realize these goals we need to solve some challenging problems
in signal processing and inference, such as finding rare and weak signals that are buried in non-stationary and
non-Gaussian instrument noise, dealing with high-dimensional model spaces, and locating what are often
extremely tight concentrations of posterior mass within the prior volume. Gravitational wave detection
using space based detectors and Pulsar Timing Arrays bring with them the additional challenge of having
to isolate individual signals that overlap one another in both time and frequency. Promising solutions to
these problems will be discussed, along with some of the challenges that remain.

\end{abstract}

\section{Introduction}

According to Einstein's General Theory of Relativity, the merger of two black holes should
shine brighter in gravitational waves than the light from all
the stars in all the galaxies in the Universe. But despite the technological advances in the century
since Einstein described his theory, we are yet to observe the most energetic phenomena in the
Universe. The first direct detection of gravitational
waves will likely come no later than the centennial anniversary of Einstein's 1916 paper on
general relativity (Einstein 1916), or at least by the anniversary of his 1918 paper on gravitational
waves where he corrected several errors in the original description of the phenomena (Einstein 1918).
The reason for this optimism is that the kilometer scale LIGO (Laser Interferometer Gravitational Observatory)
instruments in the US (Abbott et al. 2009a) and the Virgo instrument in Italy (Acernese et al. 2008)
will soon be back on line after undergoing major upgrades that are designed to increase their
sensitivity by a factor of 10, and the volume of space explored by a factor of 1000 (Harry 2010, Accadia
et al. 2011). Even with fairly conservative estimates for the astrophysical event rates (Abadie et al. 2010),
we should be seeing multiple detections each year once the detectors reach design sensitivity.

The great French mathematician Pierre-Simon Laplace anticipated several of the key elements in
gravitational wave astronomy, most notably the gravitational collapse of stars to form black
holes (Laplace 1796),
and the effects on a binary orbit due to a finite propagation speed for the gravitational force -- though
the specific form for the force law that he considered violates Lorentz invariance and leads to the
prediction that binary systems will absorb rather than emit gravitational waves (Laplace 1805).
But Laplace would likely be puzzled (and perhaps a little dismayed) by the way the data are currently
being analyzed, where talk abounds of ``detection statistics'' and ``false alarm and false dismissal
probabilities''.
There are several reasons why this frequentist approach to signal analysis has held sway in gravitational
wave astronomy. Some can be traced to the historical development of the field, and some are related to
the computational challenges of implementing the Bayes-Laplace approach to inference. However, I will argue
that the main reason that the frequentist approach continues to prevail can be traced to the difficulty in
specifying a suitable form for the likelihood, combined with an inflexible application of the Bayesian approach.
Just as Monte Carlo studies are used to tune and calibrate the frequentist detection statistics, we need
to take a more experimental approach to defining the likelihood. Over the past few years there has been a
significant increase in the number of papers applying Bayesian inference to gravitational wave astronomy, and
it is possible to sense a sea change in the community. In my description of gravitational wave signal analysis
I will follow a Bayesian approach, with occasional diversions to connect with the traditional description in
terms of matched filtering, chi-square statistics and signal-to-noise ratios.

I will begin with a brief description of gravitational waves and methods for their detection, and outline
some of the questions we hope to answer with these observations. In describing the signal analysis I will
endeavor to use a minimum of jargon in the hope that the approaches we are using may prove useful elsewhere,
and in the hope that practitioners from other fields can suggest useful techniques that we may not be aware of.
At a fundamental level, all data analysis problems are the same, and while the parts of a particular analysis
that prove the most challenging will differ from problem to problem, the same sticking points have a habit of
showing up in many different fields of study. 

\section{Gravitational Wave Astronomy}

The key properties of gravitational waves can be derived from the linearized Einstein equations.
In this description the spacetime metric is decomposed into a slowly varying background $g_{\mu\nu}$ and a small,
rapidly varying perturbation $h_{\mu\nu} \ll 1$ (Isaacson 1968). In what follows I will suppress the tensorial indices
on the dimensionless gravitational perturbation $h$. The linearized Einstein equations tell us that gravitational
waves travel at the speed of light and come in two transverse polarizations states. The physical nature of the
waves can been seen by computing the Riemann curvature tensor from the perturbed metric, which reveals
that gravitational waves manifest as a time-varying, quadrupolar tidal field that alternately squeezes then
stretches space along orthogonal directions that lie in the plane perpendicular to the direction of propagation.
The fractional change $\Delta L/L$ in the distance between two free masses is proportional to the gravitational wave
strain $h$ up to geometrical factors of order unity (at least in the limit where $L$ is small compared to
the gravitational wavelength). The two polarization states produce distortion patterns that are rotated by
$45^\circ$ with respect to each other, and are referred to as the ``plus'' $+$ and ``cross'' $\times$ polarizations.
Conservation of mass and linear momentum forbid the emission of monopole or dipole radiation, so to leading order it
is the time varying quadrupole moment of a mass distribution that is responsible for gravitational wave
generation (Einstein 1918).

Simple back-of-the-envelope calculations can be used to arrive at surprisingly accurate estimates for the
strength, frequency, luminosity and duration of gravitational wave signals for self-gravitating systems.
The quantities that enter these expressions are the internal and external gravitational field strengths
and the light crossing time of the source. The dimensionless gravitational potential is given by
\begin{equation}
\Phi = \frac{G M}{c^2 r} \, ,
\end{equation}
where $M$ is the mass-energy involved in generating the gravitational waves and $r$ is either the distance to
the source $D$ (for the external potential) or the size of the source $R$ (for the internal potential). The light
crossing time is simply $T=R/c$. The waves have amplitude
\begin{equation}
h \sim \left( \frac{G M}{c^2 D} \right)\left( \frac{G M}{c^2 R} \right),
\end{equation}
frequency
\begin{equation}\label{kepler}
f \sim \frac{1}{T} \left( \frac{G M}{c^2 R} \right)^{1/2},
\end{equation}
evolution time scale
\begin{equation}
\frac{f}{\dot f} \sim T \left( \frac{G M}{c^2 R} \right)^{-3},
\end{equation}
and luminosity
\begin{equation}
{\cal L } \sim \frac{G}{c^5} \left( \frac{G M}{c^2 R} \right)^{5} .
\end{equation}
The quantity ${\cal L}_* = G/c^5 = 3.6 \times 10^{59} {\rm ergs}\; {\rm s}^{-1}$ sets the
maximum luminosity of a gravitational wave source in general relativity as the potential
$G M /c^2 R$ saturates at around unity for a black hole. For binary systems equation (\ref{kepler}) is
just Kepler's third law in disguise. Plugging in some numbers for a source that may occur with a reasonable
event rate, such as the merger of a binary system of stellar remnant black holes at the distance of the Coma
cluster, we find that the amplitude of the waves reaching the Earth is a paltry $h \sim 10^{-22}$ despite an
energy release of over $10^{54}$ ergs in the final second of the merger.

The small amplitude of the waves can be attributed to the extreme stiffness of spacetime.
Drawing an analogy between Einstein's equations and Hooke's law for elastic solids implies that spacetime has a very
large ``metrical elasticity'' $c^4/8 \pi G$ (Sakharov 1968), and a correspondingly large specific
impedance $c^3/G = 4.5 \times 10^{38} \, {\rm g}\; {\rm s}^{-1}$. This is vastly larger than the specific impedance of
materials such as aluminum $\sim 10^{10} \, {\rm g}\; {\rm s}^{-1}$ which are used to make resonant bar detectors. The
difficulty in detecting gravitational waves with bar detectors can be seen as an impedance matching problem. In
contrast, the dense nuclear material that makes up a Neutron star has a specific impedance of around
$10^{36} \, {\rm g}\; {\rm s}^{-1}$, making the vibration modes of these objects promising sources of
gravitational waves.

While the weak coupling of gravitational waves to matter poses a challenge for detection, it also makes them
the ultimate form of ``x-ray vision'' for seeing deep inside the most extreme environments in the Universe.
When the iron core of a massive star collapses in the first stage of a supernovae explosion, the
density is so great that even neutrinos are trapped and the only signals that are able to escape and provide
information about the dynamics are gravitational waves. Gravitational waves are also the only signals that can
reach us from the first fraction of a second after the big bang -- electromagnetic signals cannot penetrate
beyond the recombination era some 400,000 years after the big bang.

\begin{figure}
\begin{center}
\includegraphics[width=5in]{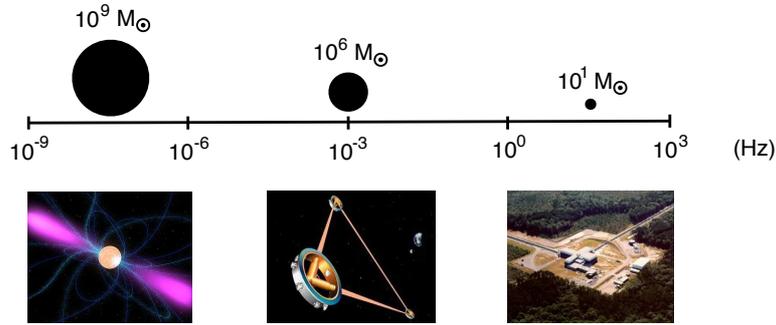}
\end{center}
\caption{The gravitational wave spectrum covered by precision pulsar timing, space based interferometers
and terrestrial interferometers. The masses of the black hole binaries to which these detectors are most
sensitive are indicated in units of solar mass.}
\end{figure}

There are two basic approaches to detecting gravitational waves. The first to be developed were acoustic transducers
that convert the strain exerted by the time varying tidal gravitational field
into the vibrations of a mechanical system. Modern versions of these detectors achieve impressive sensitivities,
but in a limited frequency band set by the natural vibrational frequency of the detector. The second approach
looks for changes in the light travel time between two free or partially constrained masses. One novel implementation
of this approach uses the radio pulses from rapidly rotating Neutron stars as a very stable time reference, and
looks for small variations in the arrival times as a signature of gravitational waves imparting a time delay or
advance. It has been predicted that binaries composed of supermassive black holes with masses $10^7$
to $10^9$ times larger than our Sun, orbiting with periods on the order of years would impart periodic variations
in the pulse arrival time of order a few nanoseconds. Using a large collection of millisecond pulsars to form
a Pulsar Timing Array, this level of variation should be detectable within the next decade (Hobbs et al. 2010).
The same idea can be implemented on a smaller scale using laser beams as a time reference. The kilometer scale ground based LIGO
and Virgo detectors operate in this way, as will the much larger space based interferometers that are currently
being planned. The LIGO and Virgo detectors
demonstrated incredible levels of sensitivity in their initial science runs, operating at strain
sensitivities of $h \sim 10^{-21}$. This equates to tracking the center of mass motion of the mirrors to less
than one-thousandth the width of a proton.
In contrast to acoustic detectors, these light-travel time detectors have broad band sensitivity
that spans several decades in frequency. Between them, these detectors cover the frequency band between $10^{-9}$
and $10^3$ Hz, as shown graphically in Figure 1. Regardless of the detection method, the detector output is
a time series that encodes the time-varying gravitational wave strain $h(t)$. 

The goals of gravitational wave astronomy extend far beyond making the first direct detection, and we are
already seeing important astrophysical results coming from the bounds that can be derived from non-detections (Abbott et al. 2008, 2009b, 2010).
There are currently large uncertainties about the event rates for black hole and neutron star mergers (Abadie et al. 2010), and
establishing these rates will tell us a great deal about the poorly understood processes by which these systems
form. If we are lucky enough to record a nearby core-collapse supernova explosion the insight into the internal
dynamics of the explosion mechanism will be a boon for efforts to numerically simulate these events. The list
of applications goes on and on, but if history is any guide, many of the most important results probably have
not been anticipated.

In addition to providing a powerful new tool for astronomers, gravitational wave detectors can also be used to
perform precision tests of Einstein's theory of gravity. At present, there are many basic predictions that await
testing, such as gravitational waves traveling at the speed of light and coming in two transverse polarization
states. In alternative theories of gravity the waves can travel at different speeds, and there are as many as
six polarizations states, some of which are longitudinal. Existing tests probe the static, weak field regime
with gravitational potentials in the range $\Phi \sim 10^{-8} - 10^{-6}$ and orbital velocities in the range
$v/c \sim 10^{-4} - 10^{-3}$. It is easy to construct candidate theories that are consistent with all existing
data, and yet differ significantly from general relativity in the final stages of a black hole merger where
$\Phi$ and $v/c$ approach unity (Yunes and Pretorius 2009).

\section{Gravitational Wave Signal Analysis}

The read-out from a gravitational wave detector is a time series of some quantity that can be related to the
gravitational wave strain. Common examples of the quantity being measured include phase differences,
frequency shifts, time delays and displacements. For LIGO the response
is derived from an error point in the differential arm length servo control loop. In any case, the output
can be converted into a time series for the dimensionless strain $s(t)$ registered by the detector, which
includes contributions from a gravitational wave signal $h(t)$ and instrument noise $n(t)$:
\begin{equation}
s(t) = R(t,\tau)\star h(\tau) + n(t) \, .
\end{equation}
The detector response function $R(t,\tau)$ depends on the polarization angle and sky location, and
includes contributions from the different times $\tau$ that the wavefronts encounter the reference masses
in the detector. These delays are important for space based detectors and Pulsar Timing Arrays, but are
generally negligible for individual ground based detectors as the light crossing time for the detector
is much less than the period of the waves they seek to detect. The delays are important for
a world-wide network of gravitational wave detectors, where we use a standard reference for $t$, such as
Greenwich Mean Time, and the times $\tau$ record when the waves arrive at each detector. With three or
more detectors in the array it is possible to triangulate where the signals came from. Additional directional
information is also encoded in the antenna patterns.

Very often we have multiple data streams $s_i(t)$, which can be stacked together to form a single data vector:
\begin{equation}
{\bf s}= {\bf R} \cdot {\bf h} + {\bf n}
\end{equation}
The individual elements of the data vector are then the time samples in a particular detector. Alternatively,
we may choose to transform to some alternative representation where the
elements might be Fourier or wavelet amplitudes. The response matrix ${\bf R}$ is a known function that can
be computed for any polarization and sky location. Our task is to infer the properties of the gravitational
wave signal ${\bf h}$ from the data ${\bf s}$, and to asses our degree of confidence in the detection by
comparing the evidence of models that include gravitational wave signals with models that only
describe the instrument noise. In a Bayesian approach to this inference problem we need to fully specify
models for the gravitational wave signals and the instrument noise. We also need to provide
priors on the quantities that enter these models, and define a likelihood function. Once this is done we can
attack the purely mechanical task of computing the posterior distribution for our model parameters and the
model evidence using techniques such as Markov Chain Monte Carlo and Nested Sampling. A second level of analysis
can then follow that uses the information gleaned about individual systems to constrain astrophysical models
for the sources (e.g. the mechanisms at play in a core collapse supernova explosion), and models
for the source populations (e.g. which stellar evolution pathways lead to the formation of neutron star or
black hole binaries).

\section{Likelihood function and models for the instrument noise}

There are two basic approaches to the problem of extracting information about the gravitational wave signals.
The first of these works directly with the raw data, while the second deals with quantities that can be
derived from the raw data. In either case, it is necessary to define a likelihood function, and this turns out
to be a surprisingly challenging task. In a recent article Skilling writes ``The instrument acquiring the data
can usually be calibrated with known inputs to find out how often it produces specific outputs, which effectively
fixes the likelihood to any desired precision. If there remain any unknown calibration parameters in the likelihood,
they can be incorporated as extra parameters to be determined, leading to extra computation but no difficulty in
principle'' (Skilling 2010). While fine in theory, the calibration of the likelihood turns out to be extremely
challenging in practice for gravitational wave detectors such as LIGO and Virgo. There are literally dozens of people devoted
to characterizing and calibrating these instruments. The difficulty is that the signals we
are looking for are rare and weak, so we need to be able to map the likelihood far into the tail of the
distribution. Compounding the problem is the fact that the behavior of the detectors changes over time due to
environmental changes and modifications that are made during the weekly maintenance activities.

To understand the challenges involved in defining the likelihood, consider a direct approach that works
with the raw data. Here we need a model $M$ for the gravitational wave signals ${\bf h}$ that may be
present in the data. We can then subtract this model from the data to form up the residual ${\bf r}
= {\bf s} - {\bf R}\cdot {\bf h}$. If our model matches the gravitational wave signal contained
in the data then the residual should be consistent with pure instrument noise ${\bf n}$. Thus the
likelihood function $p({\bf s} \vert {\bf h}, M)$ is nothing other than the joint noise distribution
$p_n({\bf n})$:
\begin{equation}
p({\bf s} \vert {\bf h}, M) = p_n({\bf r}) \, .
\end{equation}
In other words, the likelihood is the noise model. A typical stretch of data from one of the
LIGO/Virgo instruments can be approximately modeled as stationary, colored and normally distributed noise
defined by the expectation values
\begin{eqnarray}
&& E[n(f)] = 0 \nonumber \\
&& E[n(f) n^*(f')] = \frac{T}{2} \delta_{f f'} S_n(f) \, ,
\end{eqnarray}
where $S_n(f)$ is the one-sided noise spectral density at frequency $f$ and $T$ is the observation
time. The joint noise distribution in this case factors into a product, and the likelihood becomes
\begin{eqnarray}
p({\bf s} \vert {\bf h}, M) &=& \prod_f \frac{1}{2 \pi T S_n(f)}
\exp\left( -\frac{r(f) r^*(f)}{T S_n(f)} \right)\nonumber \\
&= & C e^{-(r\vert r)/2} = C e^{-\chi^2/2}\, ,
\end{eqnarray}
where in the last line the product of the exponentials has been converted to a sum of exponents to define
the noise weighted inner product
\begin{equation}
( a \vert b) = \frac{2}{T} \sum_f \frac{a(f)b^*(f) +a^*(f)b(f)}{S_n(f)} \, .
\end{equation}
This is nothing other than the standard chi-squared goodness of fit first introduced by Gauss. 
The likelihood for a network of detectors
with independent noise realizations is simply the product of the individual likelihoods. If correlations
are present they can be dealt with by introducing a noise correlation matrix.

In the frequentist approach, the Neyman-Pearson criterion
is used to identify detection candidates by minimizing the false
dismissal probability for a given false alarm probability. In stationary, Gaussian noise the
optimal decision statistic for the Neyman-Pearson test is the likelihood ratio
\begin{equation}
\Lambda = \frac{p({\bf s} \vert {\bf h}, M)}{p({\bf s} \vert 0, M)} = \exp\left(-({\bf s}\vert {\bf R}\cdot{\bf h})
+\frac{1}{2}({\bf R}\cdot{\bf h}\vert {\bf R}\cdot{\bf h})\right).
\end{equation}
The space of possible observations is partitioned by setting a threshold $\Lambda_0$ that yields the desired
false alarm probability. The threshold can be computed analytically for the simple noise model considered here.
Maximizing $\ln \Lambda$ with respect to the amplitude of the signal
$A = ({\bf R}\cdot{\bf h}\vert {\bf R}\cdot{\bf h})^{1/2}$ yields the detection statistic
\begin{equation}
\rho = \frac{ ({\bf s}\vert {\bf R}\cdot {\bf h})}{({\bf R}\cdot{\bf h}\vert {\bf R}\cdot{\bf h})^{1/2}} \, .
\end{equation}
The maximum value of $\rho$ is referred to as the amplitude signal-to-noise ratio (SNR) of the best fit
signal ${\bf h}$. The SNR can be computed on a per detector basis, and for the full detector network.
The quantity $\rho$ is referred to as the Wiener matched filter statistic, and it can be shown to be the
optimal statistic in stationary Gaussian noise for signals with known amplitude. It is the reason why
almost every book, talk and paper on gravitational wave astronomy makes frequent references to ``matched
filtering'', and cites the power of this technique in lifting signals that are buried beneath the instrument
noise. Unfortunately, none of the assumptions used to motivate this approach pertain in practice, and
the real world analysis is far more complex.

\begin{figure}
\begin{center}
\includegraphics[width=4in]{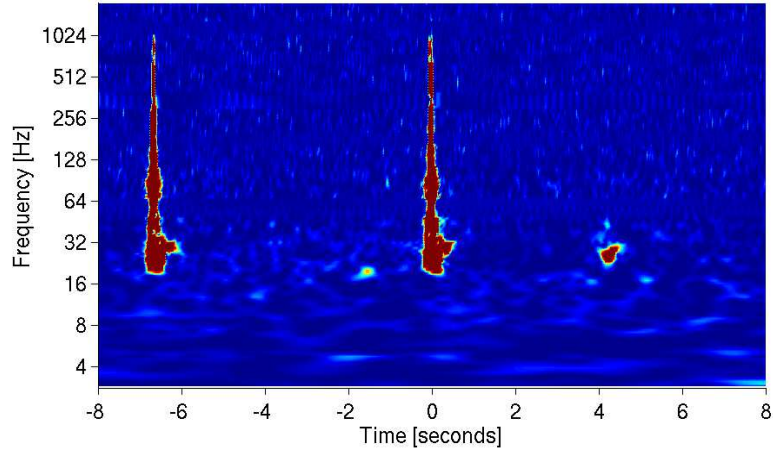}
\end{center}
\caption{A scalogram showing examples of non-stationary features in data from one of the LIGO detectors (from Slutsky et al. 2010).}
\end{figure}

% http://arxiv.org/abs/1004.0998

Figure 2 shows a time-frequency map of a small segment of data from the LIGO detector in Hanford Washington,
where a loud transient feature or ``glitch'' stands out above the surrounding noise. Studies have shown that
these instrument glitches follow a power law distribution in amplitude, with a greater prevalence of
low amplitude glitches. In relatively clean stretches of data with no obvious glitches the noise looks
Gaussian, but sections with glitches develop heavier tails. It is clear, however, that glitches do more than
fatten the tails of the distribution, they also introduce correlations across the time-frequency plane.
Ideally we would like to follow Skilling's advice and fully characterize this more complicated noise model to
arrive at a realistic description of the likelihood. Short of this ideal, I will describe some of the methods
that are being used to account for the non-Gaussianity and non-stationarity of the data.

In the traditional approach, statistics like $\rho$ and $\chi^2$ are adopted, and rather than using
theoretical distributions for these quantities, Monte Carlo techniques are used to estimate the probability
distributions in pure instrument noise and when signals are present. Since we do not know in advance if a
given segment of data contains a gravitational wave, unphysical time shifts are introduced between
different instruments. This ensures that any signals that might be present will fail the coincidence test
that is used to decide if the signals seen in each detector are consistent within the
light travel times between the detectors. The behavior when signals are present is determined by adding simulated
signals to the data. The analysis also makes use of the thousands of monitors attached to the detectors, such as
seismometers, microphones and ammeters. Sections of data where correlations are detected between
the gravitational wave channel and the environmental channels are vetoed and discarded.

In recent years the traditional approach has been extended to include a large number of quantities that
can be measured from the raw data. These include the parameters $\vec{\lambda}$ that define a particular
gravitational wave model ${\bf h}(\vec{\lambda})$ that yields the highest SNR in each detector, as well
as the individual SNRs and the standard $\rho$, $\Lambda$ and $\chi^2$ statistics. The distributions of
these parameters for the signal-plus-noise model and noise-only model are then estimated using the time slide and
signal injection procedure described above. The samples are then fed into a multi-dimensional classifier
to produce a decision tree that assigns a probability that an event is consistent with noise (Hodge 2011), or are used
to develop joint likelihoods that feed into a Bayesian model selection analysis (Cannon 2008). Both of these approaches
are better able to account for the non-Gaussianity of the instrument noise than the traditional approach.

More recently, a new approach for working with the raw data has been put forward that includes a sophisticated
noise model that explicitly accounts for glitches in the data by introducing a glitch model ${\bf g}$ (Littenberg and Cornish 2010).
The noise is then split into two components: ${\bf n} = {\bf n}_R + {\bf g}$, where the glitches are modeled as localized
concentrations of energy in the time-frequency plane, while the ``regular'' component ${\bf n}_R$ is assumed to
be uncorrelated in frequency and well approximated by a Gaussian or some other similar simple distribution.
The {\em BayesWave} algorithm for implementing this approach will be described in a subsequent section.

\section{Signal Models}

Typical analyses focus on a single class of gravitational wave signals. Examples include the inspiral signal from
a Neutron Star binary, the merger signal from two colliding black holes, and stochastic signals from processes
in the early Universe. The more tightly focused the search, and the more prescriptive the signal model, the deeper
we can dig into the noise. On the other hand, we also want to be open to finding unexpected signals about which
little is known.

Within the confines of general relativity, the most general signal model is one in which the waves come in two
transverse polarizations and travel at the speed of light. The parameters in this model are the discrete time
samples $h_\times(t)$ and $h_+(t)$. Using three or more independent and non-aligned detectors it is possible to
infer the presence of bright gravitational wave signals that stand out above the noise in each
detector (Klimenko et al. 2005, Chatterji et al. 2006). In order
to detect weaker signals we need more restrictive models for the signals and the the instrument noise. For example,
we could demand a degree of smoothness in the signal model by parameterizing the signal in terms of control points
of a cubic spline. This reduces the number of parameters in the signal model, and allows us to find signals that
are slightly below the noise level. Similar gains can be achieved by adopting a time-frequency decomposition of the
data such as a discrete wavelet basis and demanding that the signal is clustered in time and frequency. The model
can be further narrowed by imposing restrictions on the frequency range and development in time. For example, the
signal from a black hole binary should increase in amplitude and frequency as the system nears merger, so we can
narrow our signal model to focus on chirp-like signals. Following this line of development we eventually arrive
at the so-called template-based analyses where approximate solutions to Einstein's equations are used to predict
the waveforms that are produced by various types of astrophysical sources. Rather than using discrete time samples
or wavelet amplitudes, these models describe the signals $h(t,\vec{\lambda})$ in terms of a relatively
small number of physical parameters $\vec{\lambda}$ that describe the source. One of the most important examples
are the templates used to describe the inspiral and merger of two black holes, which are derived using a combination of
analytic and numerical approximations, and yield waveforms that depend on just 17 parameters. The list of parameters
includes the masses and spins on the black holes, the distance from the detector, and the sky location.

Template-based searches allow us to detect signals that are buried in the instrument noise. The explanation for
this is usually given in terms of matched filtering, and is attributed to a coherent build-up of power
over many wave cycles in the cross-correlation of the signal and the template. While coherent integration over
many cycles is important, the same coherent build-up occurs in models without templates (Klimenko et al. 2005).
The real reason why template based models outperform less prescriptive signal
models is that they involve far fewer parameters, and thus yield much higher evidence values. A template
based search for the inspiral signal from a binary neutron star may detect signals with integrated signal-to-noise
${\rm SNR} \sim 7$ that builds up over $N \sim 10,000$ cycles. This corresponds to an average per-cycle or raw
signal-to-noise ${\rm S/N}= {\rm SNR}/\sqrt{N}$ of around 0.07. In contrast, the most general
signal model described above requires ${\rm S/N} > 1$, and integrated signal-to-noise ${\rm SNR} > 100$.

Another advantage of template-based models is that they allow us to extract detailed information about the
sources. Posterior distributions for $p({\bf h} \vert {\bf s})$ become posterior distributions for the
model parameters $p(\vec{\lambda} \vert {\bf s})$, and we can quote confidence intervals for
quantities such as the masses, distance to the source, and the sky location. The latter is very important for
performing coordinate observations with electromagnetic telescopes. Connecting with physical parameters
is more difficult in models that work directly with the waveforms, but it is possible to produce posterior
distributions for quantities such as the duration, peak frequency, time-frequency volume and degree of
polarization.

Rather than trying to infer $h(t)$ directly from the data, some analysis techniques use cross-correlations
between data streams. The idea is that the signal components combine coherently while the instrument noise
contributions average to zero. This approach is used to search for stochastic signals and long duration signals
of unknown morphology (Allen and Romano 1999). By introducing appropriate time delays in the cross-correlation,
the data from widely separated instruments can be used to perform a directional search (Mitra et al. 2008,
Dhurandhar et al. 2008).

\section{Detection and Characterization}

The question of whether a signal has been detected or not is central to gravitational wave astronomy. In a Bayesian
setting we can attempt to answer this question by computing the odds ratio between competing hypotheses.
Even without a good detection candidate it is possible to constrain event rates and the astrophysical
models used to predict these rates. 

Physicists have a natural attraction to the idea that there is a number we can calculate that tells us which
model better describes the data. It is easy to forget that the output of this calculation is only as good
as its inputs. The choice of likelihood function can have a dramatic effect on the evidence and the odds ratio. What
might seem like an extreme outlier when using a Gaussian distribution may be unsurprising using a Cauchy distribution. It
is also important to be clear about what hypotheses are being tested. We are not making statements about wether
gravitational wave signals have been detected, but rather, we are comparing some simplified noise model to some
subset of possible signal models. 

The detection problem is further complicated by the ``needle-in-a-haystack'' challenge of locating weak signals in
a vast search space. For example, to yield a good fit to a binary black hole merger signal in LIGO/Virgo data
requires a template with a merger time that matches that of the signal to within a few milliseconds in a data set
that spans tens of billions of milliseconds. Other quantities, such as the ``chirp mass'' (define by ${\cal M}
= (m_1 m_2)^{3/5}/(m_1+m_2)^{1/5}$, where $m_1$ and $m_2$ are the individual component masses) also need to
match to within tiny fractions of their prior range. Put another way, the region $\Delta V$ that contains 99\% of
the posterior mass occupies a minute fraction of the total prior volume $V$. Ratios in excess of $10^{40}$ are
not unheard of. Standard implementations of Markov Chain Monte Carlo (MCMC) and Nested Sampling algorithms have no
hope of finding these regions. Instead, the analysis proceeds in two steps, starting with a search phase that
attempts to locate the modes of the posterior, followed by a characterization phase where the regions of high
posterior weight are mapped and the evidence is calculated.

During the search phase anything goes, and we are free to maximize over the parameters used to describe the
templates. The time of arrival can be maximized using a standard Fourier transform trick, as can the initial phase.
The amplitude or distance to the source can be solved for by analytically maximizing the posterior. In some cases
it is possible to maximize over several more parameters. These maximization techniques can dramatically shrink the
search space. When few parameters remain to be searched over it becomes computationally feasible to methodically
work through a grid of templates that have been layed out so as to ensure some minimum overlap for signals with
parameters that lie between the grid points. Most of the current LIGO/Virgo searches are performed using template
grids. When the remaining search space has more than three or four dimensions it becomes necessary to use other
techniques, such as hierarchical grids or semi-stochastic searches based on genetic algorithms or greedy non-Markovian
variants of MCMC algorithms (Cornish 2011). The posterior distributions for these more complex waveforms are often multi-modal,
narrowly concentrated and highly curved when plotted in terms of the physical parameters that define the waveforms.
The efficiency of the search algorithms is highly dependent on the parameterization that is used, and physical insight
can help guide the selection of parameterizations where the posterior distribution takes a simpler form. In many
cases it is possible to use the functional form of the templates to predict where the secondary modes will be located.
All it takes is for the search to find a secondary, tertiary or even a quaternary mode for the full posterior structure
to be revealed (Cornish 2011).

The next task is to map out the posterior distribution and compute the model evidence. This is done using a combination
of MCMC and Nested Sampling algorithms. Multi-modality and strong correlations between parameters make these
distributions difficult to explore. This has lead to the development of what can best be described as ``everything
but the kitchen sink'' MCMC algorithms that combine a wide variety of techniques, such as parallel tempering,
differential evolution, delayed rejection, mode hopping proposals and local diagonalization of the covariance matrix. 
The latter technique can be applied to samples taken from the history of the chain (a procedure that can be shown to
be asymptotically Markovian), or by computing the Fisher information matrix. Parallel tempering is very effective in
helping the chains move between local modes since the hot chains with ``temperature'' $T > 1$ explore a flatter
likelihood landscape with $p({\bf s} \vert {\bf h})_T = p({\bf s} \vert {\bf h})^{1/T}$. As an added bonus, having
chains at different temperatures allows the the evidence to be computed via thermodynamic integration, though attaining
accurate estimates using this procedure can be computationally expensive. Nested sampling has become the preferred
method for computing the evidence, especially since the general purpose MultiNest software package has been made publicly
available (Feroz, Hobson and Bridges 2009). As with the MCMC routines, MultiNest needs help to locate the tightly
concentrated modes from within
the vast prior volume. One solution is to run MultiNest on small sub-volumes that contain the vast majority of the
posterior mass and sum the contributions to the evidence. These sub-volumes can be identified from the MCMC chains.
To ensue that the majority of the posterior mass is enclosed it is a good idea to use the hot ($T\sim 4 \rightarrow 9$)
MCMC chains, as these explore regions of lower likelihood.

\section{Likelihood Redux}

The posterior distributions and odds ratios are highly dependent on the choice of the likelihood function.
As Sivia
and Rawlings have pointed out, people are ``led to regard the prior as subjective, whereas the likelihood
is seen as objective. In reality, each is just a conditional PDF and, as such, they are on par with each
other'' (Sivia and Rawlings 2010). At least a poor choice of prior can be overwhelmed by sufficiently informative
data, whereas the bias introduced by a poor choice of likelihood grows with the information content.

But how do we know if we have made a bad choice for the likelihood function? One way is to look for
systematic bias in parameter recovery using an ensemble of signal injections. For example, do the injected
signal parameters lie within the 90\% confidence intervals 90\% of the time? This frequentist style of
testing can also be used to compare Monte Carlo studies of false positive rates to what would be expected based
on the odds ratio. It has even been suggested that we simply accept that our likelihood function is wrong, and
treat the Bayes factors as a frequentist detection statistic to be calibrated using time slides and signal
injections (Veitch and Vecchio 2008). But this does not address the issue of bias in the posterior distributions (Raymond 2010).

Several authors have put forward likelihood functions that allow for non-Gaussianity and non-stationarity in
the data (Principe and Pinto 2009, Allen et al. 2003). The {\em BayesWave} algorithm introduced earlier
takes a different approach, and rather than
building in the effect of instrument glitches, it seeks to model individual glitches and remove them from the
data (Littenberg and Cornish 2010). The glitch model ${\bf g}$ uses a discrete wavelet basis, with the wavelet
amplitudes as the model parameters. The
number of active (non-zero) pixels in the model is a free parameter that is determined from the data. The
likelihood function assumes an independent Gaussian distribution for the residual
${\bf r} = {\bf s} - {\bf R}\cdot{\bf h} -{\bf g}$ in each wavelet pixel. The variance of the residual in each
frequency band over a $\sim 1000$ second time interval is a free parameter to be determined from the data. Allowing
the noise level to float in this way lets us model slow drifts in the instrument sensitivity over time. The
{\em BayesWave} algorithm is implemented using a Reversible-Jump MCMC (RJMCMC) routine that allows the number of active
glitch pixels to vary. Relatively quiet stretches of data light up very few glitch pixels, while loud instrument
glitches can light up hundreds of glitch pixels. The signal model can use waveform templates for targeted searches
or wavelet amplitudes for a more general search. The latter does not work for a single detector as there is nothing
to distinguish between the signal and glitch models. With a network of detectors its more parsimonious to assign
coincident excess power to the signal model than the glitch model. The glitch identification can be enhanced by
using physically motivated priors, such as assigning a higher probability to glitch pixels that have active
neighbors.

The complexity of the {\em BayesWave} noise model does pose a challenge when trying to compute the model evidence.
Since the number, location and amplitude of the active pixels are all variable, we are in fact dealing with a
very large collection of distinct noise models that have to be marginalized over. Nested Sampling is ill suited to this
task, and it is more natural to extend the RJMCMC model space to include the signal model. The Bayes factor between
the noise-only and signal-plus-noise model can be estimated from the number of iterations that are spent exploring
each model. The challenge is finding proposals that allow the chains to jump between models. An un-informed jump
proposal has little or no chance of landing in the small $\Delta V$ volume where the signal posterior has significant
weight. Our solution is to do two pilot runs, one with the signal model and one without, and use the posterior
distributions from these chains as proposal distributions for the trans-dimensional jumps in the combined run with
both models (Littenberg and Cornish 2010). The distributions can be stored in a sparse matrix representation of a
multi-dimensional histogram with only a small number of significant cells, or in a KD data tree (Farr and Mandel 2011)
with uniform occupancy and variable cell size.

Tests have shown that the residuals produced by the {\em BayesWave} algorithm pass all the standard tests of
Gaussianity. All significant non-Gaussian features are either assigned to the signal or glitch models, and the
signal parameters show none of the systematic biases that occur with a simple Gaussian likelihood.

\section{The cocktail party problem}

\begin{figure}
\begin{center}
\includegraphics[width=4in]{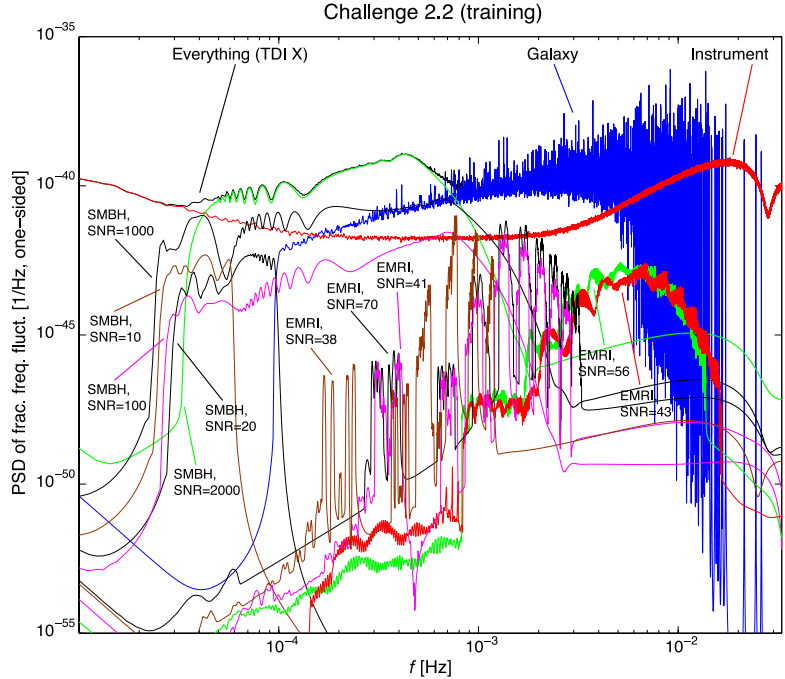}
\end{center}
\caption{A simulation of the strain spectral density recorded by a future space based gravitational wave
observatory broken out into individual contributions (Arnaud et al. 2007).}
\end{figure}

A new set of challenges awaits us in the next phase of gravitational wave astronomy,
where will have to contend with the signal-rich data streams from space-based detectors and third-generation
ground-based detectors such as the Einstein Telescope. In the near term, the data from Pulsar Timing
Arrays will pose their own challenges. The main difference from the LIGO/Virgo analysis is the
long dwell time of the signals, which results in many signals being in-band at the same time. A space-based
detector operating in the milli-Hertz regime will record the signals from tens of millions of short-period
white dwarf binaries in our galaxy, in addition to the signals from dozens of massive black hole mergers and
possibly hundreds of capture signals from stellar-mass black holes or neutron stars being swallowed by a massive
black hole (also called Extreme Mass Ratio Inspirals or EMRIs). Figure 3 shows a simulated power spectrum
for two years of data from the
proposed Laser Interferometer Space Antenna (LISA) mission that has been broken out into some of the
individual contributions. The simulation includes a complete galaxy realization,
but only a small fraction of the expected number of black hole merger and capture signals. The signals from
unresolved sources are the dominant source of ``noise'' for the LISA mission concept. From this cacophony
we need to be able to pick out and interpret individual signals, a situation akin to trying to follow a
conversation at a loud cocktail party.

The signals overlap in both time and frequency, and while having accurate template families for the signals helps,
a serial analysis is not possible due to correlations between the templates. We are faced with the challenge of
having to simultaneously extract a vast number of signals, not knowing in advance how many signals are resolvable,
or what their parameters might be. Again it is natural to separate the analysis into search and characterization
stages. The search can start with the brightest signals and gradually work down to the weakest signals while
updating the global solution along the way (Crowder and Cornish 2007, Littenberg 2011). For the characterization
stage it is natural to consider a RJMCMC implementation that generalizes the methods used to identify an unknown
number of sinusoids in noisy data (Andrieu and Doucet 1999, Umstatter et al. 2005). A realistic implementation
of this approach has yet to be demonstrated.

\section{Summary}

With the first direct detection of gravitational waves just around the corner, a large group of researchers
is racing the clock to develop robust analysis techniques to maximize the science yield from the first
detections. The application of Bayesian inference is becoming more widespread and established, but the complicated
nature of the instrumental noise is a challenging problem that still needs to be addressed. Future detectors
promise to deliver data that are so rich in signals that the unresolved components will become the dominant
source of noise. After a century of waiting for the first detection, an over-abundance of signals will be a
good problem to have.

\section*{Acknowledgments}
The author is grateful for the comments and suggestions from Albrecht Rudinger, Ray Frey, Ben Owen and
Alberto Vecchio. This work was support by NSF Award 0855407 and NASA grant NNX10AH15G.

%\section{References}

\end{document}